# Helium recovery system at IB3A

**D. Porwisiak[1,2]\*, M.J. White[1], B.J. Hansen[1]**

[1] Applied Physics and Superconducting Technology Directorate, Fermi National Accelerator Laboratory, Batavia Il, USA
[2] Faculty of Mechanical and Power Engineering, Wroclaw University of Science and Technology, Wroclaw, Poland

\*E-mail: dporwisi@fnal.gov

**Abstract.** The growing demand for sustainable cryogenic operations at Fermilab has underscored the need to improve helium management, particularly at the Industrial Building 3A (IB3A) test facility. IB3A characterizes and tests superconductors, cables, and coils for projects such as the HL-LHC AUP and Mu2e, yet currently relies on 500 L Dewars whose boil-off is vented to atmosphere, wasting a critical, non-renewable resource and increasing the cost of testing. A project is therefore under way to link IB3A to an existing purification and liquefaction station in a neighboring building through a dedicated pipeline. Captured helium will be transferred, purified, reliquefied, and returned for reuse, cutting losses and operating costs. This paper details the first two project phases: "Design and Engineering" and "Procurement and Installation." The design phase finalizes pipeline specifications, establishes flow-control requirements, and resolves integration challenges with existing cryogenic infrastructure. The procurement and installation phase covers material sourcing, pipeline construction, and deployment of control and monitoring systems to assure reliable, efficient operation. Key technical hurdles—route optimization, pressure-drop mitigation, and interface compatibility—are discussed alongside implemented solutions. Implementing the pipeline and upgrading IB3A will dramatically reduce helium consumption and therefore testing cost, strengthening Fermilab's capacity to support frontier science far into the future.

1. Introduction

An assessment of Fermilab's Industrial Building 3A (IB3A) cryogenic test facility—where superconductors, cables and coils are characterized for internal R&D and major programs such as the US High-Luminosity LHC Accelerator Upgrade, Mu2e and various external requests—has revealed an urgent need for a closed-loop helium recovery system. Historically, the IB3A facility relies on 500 L dewars and simply vents all helium boil-off to atmosphere, resulting in substantial annual losses presented in Table 1. By installing the IB3A recovery piping, Fermilab will dramatically reduce its consumption of this non-renewable resource and lower operating costs.

**Table 1.** IB3A Helium consumption

| Fiscal year | Helium consumed, L | Cost, $k |
|---|---|---|
| FY23 | 7,500 | $135 |
| FY22 | 17,956 | $125 |
| FY21 | 46,763 | $325 |
| FY20 | 25,185 | $170 |
| FY19 | 33,438 | $213 |

## 2. Project objectives

The primary objective of this project is to implement a helium recovery system by capturing gaseous helium (GHe) emitted from the four Teslatron cryostats housed in the IB3A test facility and routing it directly to the purifier compressor suction at IB1 (Industrial Building 1) (See Figure 1.). By doing so, we will enable the purified helium to be reintroduced into IB1's test facility operations, effectively reducing the need for external helium procurement for this facility. IB1 cryogenic system is further described in [1]. This approach not only maximizes resource efficiency and minimizes environmental impact but also generates substantial cost savings by offsetting the expenses associated with purchasing high-purity helium for routine facility use. A future phase of the project may include adding a dewar liquid fill station at IB1.

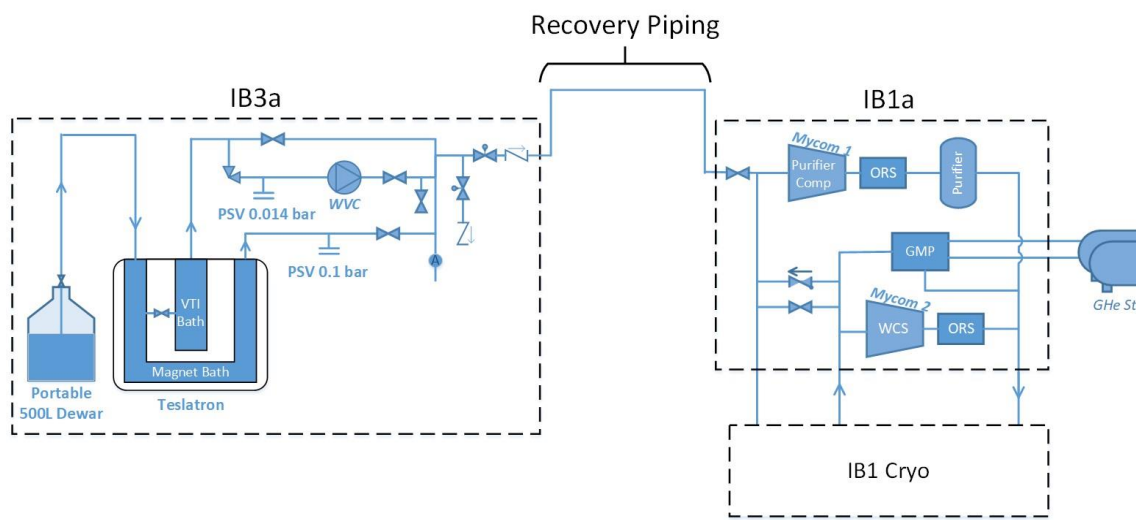

**Figure 1.** Flow diagram of IB3a helium recovery. WVC stands for vacuum pump, PSV is pressure safety valve, A is $N_2$ Analyser.

### 3. Design and engineering

Helium from the IB3A test facility will be recovered via over the roof piping which will connect IB3A Teslatron cryostats (Figure 2A) with the IB1A purifier compressor system. The pipe consists of 2 main sections – DN100 (4" NPS) (245 m length starting at IB3A, ending on Midway roof between ICB and IB1A, see Figure 2B) and 3" (DN80 NPS)(38 m length starting on Midway roof, ending on IB1a). Schedule 5 pipe was used to decrease weight of the pipe. A satellite image view of industrial building complex is presented on Figure 2B.

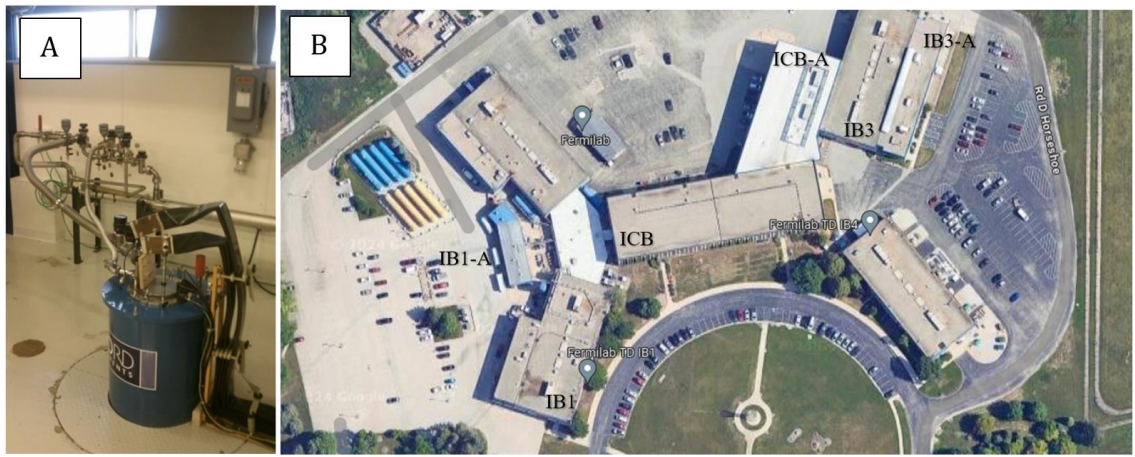

**Figure 2.** A- Teslatron cryostat, B – Top view of industrial buildings complex

The system will require minimal operational input and will allow recovery of helium from 2 cryostats at a time. Expected flow rate for normal operations (static boil-off, cooldown, warmup) will be 0.1 - 6 g/s. For operational cases, the pressure drop was estimated to be at most 15 mbar (0.21 psi). Pressure parameters are presented in Table 2. To facilitate helium recovery during 2 K operations, a vacuum pump will be installed (Rotary vane pump with roots pump booster), capable of extracting approximately 1 g/s of helium at 30 mbar. The pump will be installed in a shed next to IB3A laboratory. The system will also offer capability to recover a significant fraction of helium from controlled magnet quenches. Recovery modes are listed in Table 3. Contamination will be controlled on both sides of the system. First, before recovery starts, helium gas will flow through Servomex trace $N_2$ Analyser located in the IB3A facility. When the gas meets purity criteria flow through the $N_2$ Analyser will be closed and recovery started. Additional contamination monitoring and protection will be provided at IB1A as part of the existing purifier compressor system. Rapid rise of contamination will trigger interlock, closing flow through the recovery piping and start venting the helium gas until issue is resolved.

**Table 2.** Pressure parameters of the recovery system.

|  | Purifier suction pressure[a] | Operating pressure | Pressure drop during normal operations | Pressure drop during warmup/ cooldown |
|---|---|---|---|---|
| Design value | 1.075 bar a | 0.096 bar g | 4.8 E-4 bar g | 0.014 bar g |

[a] Can be reduced if needed

Roof piping lays on non-destructive pad supports, which allow for free horizontal movement. Maximum calculated sustained stress is 43.9 MPa (allowable 115 MPa), and maximal expansion stress is 15.8 MPa (allowable 172.7 MPa) for given operating conditions.

During the design phase, alternative recovery configurations were also evaluated, including a system that would accumulate gaseous helium in a gas bag within the IB3A laboratory and, upon reaching approximately 80 percent capacity, automatically engage a high-pressure compressor to fill a tube trailer. The captured gas could then be deployed and repurposed at the various cryogenic plants across Fermilab. However, this concept was ultimately dismissed for several reasons: the operational costs associated with transporting and emptying the tube trailer proved prohibitively high, experience from other laboratories indicated a significant risk of water contamination in the gas bag; and the limited available space to house a large gas bag in the IB3A laboratory rendered this approach impractical.

**Table 3.** Recovery modes of operation

| Operating mode | Max mass flow rate | Max allowable LHe consumption | Operating pressure | Possible scenarios |
|---|---|---|---|---|
| Cooling down/ warmup | 6 g/s | 168 L per hour | 0 - 0.1 barg | Teslatron 1 cooldown/warmup |
| Cooling down /warmup 2 | 6 g/s | 168 L per hour | 0 – 0.28 barg | Teslatron 2 or 3 or 4 cooldown/warmup Teslatron 1 normally operates and is disconnected from recovery |
| Normal operations | 2 g/s | 56 L per hour | 30 mbar a – 1.1 bar a | 1. Teslatron 1 normally operates 2. Teslatron 1 normally operates and any of the other Teslatrons also normally operates 3. Teslatron 2, 3 or 4 normally operates. |
| Controlled quench | 45 g/s | 58 L per few minutes | 0.28 psig | Only one Teslatron at a time, other cryostats isolated from recovery. Some helium will be lost through relief valves. |

**Table 4.** Project milestone schedule.

| Milestone | Date |
|---|---|
| Start Project | August 2023 |
| Final Design Complete | June 2024 |
| Installation Contract PO Awarded | Oct 2024 |
| Installation Complete | Oct 2025 |
| Safety ORC Received | Nov 2025 |
| Commissioning Complete | Jan 2026 |

### 4. Procurement and installation

High level project milestone schedule is presented in Table 4. Awarded PO for roof piping installation is already completed as shown in Figure 3. In the next phase of the helium-recovery project, the fabrication of the IB3A interface piping will commence, ensuring a seamless connection between the newly designed recovery lines and the existing roof-mounted infrastructure at the IB3A facility. Simultaneously, we are initiating the procurement process for a dedicated vacuum pump, selecting a unit whose capacity and reliability meet the requirements of continuous GHe evacuation and transfer. Once the vacuum pump is on site, our team will proceed with the installation of a comprehensive monitoring, control, and alarming system, integrating pressure, temperature, and flow sensors to provide real-time oversight of the recovery loop. Finally, we will complete the fabrication and welding of the final tie-in piping spools that link the IB3A interface directly to the IB1 purifier compressor suction, thereby closing the recovery circuit and enabling the purified helium to enter IB1 operations without interruption.

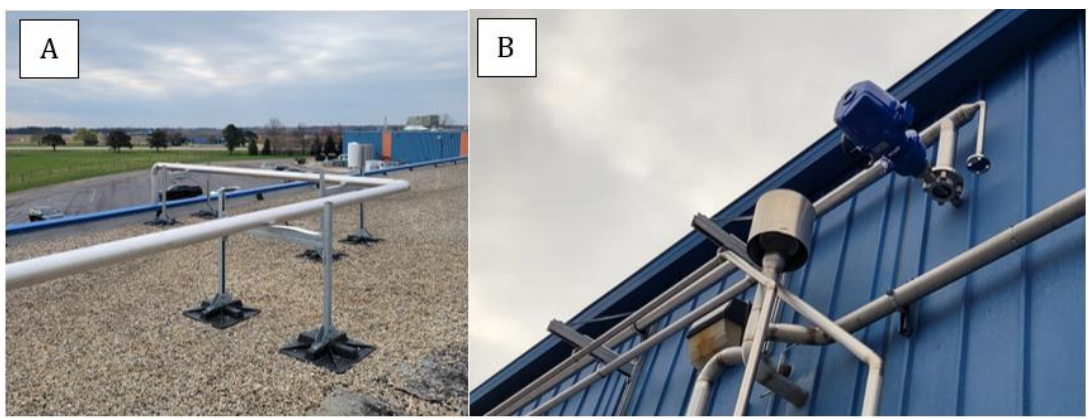

**Figure 3.** Installed roof piping A – roof at IB3A side, B – IB1a end with isolation valve.

## 5. Summary

This article describes the development and implementation of a helium-recovery system linking the four Teslatron cryostats in IB3A to the purifier compressor at IB1. After evaluating alternative configurations the project settled on direct piping from IB3A to the IB1 suction header. The primary objectives are to capture and route waste gaseous helium from IB3A into the IB1 purification loop, thereby reducing dependence on externally purchased helium. Key remaining activities include fabricating and installing an interface manifold at the IB3A roof connection, purchasing and commissioning a suitably sized vacuum pump, deploying a comprehensive monitoring, control, and alarm system, and completing final tie-spools that connect the recovery piping directly into IB1's compressor suction. Once fully operational, this integrated system will enable continuous flow of helium into IB1, generating significant cost savings and enhancing resource efficiency.

**Acknowledgments**

This work was produced by Fermi Forward Discovery Group, LLC under Contract No. 89243024CSC000002 with the U.S. Department of Energy, Office of Science, Office of High Energy Physics.